\documentclass[12pt,showpacs,showkeys,amsmath,amssymb]{revtex4}
\usepackage{amsmath,amsfonts,amsthm,amscd,amssymb,latexsym}
\usepackage{bm}
\usepackage{dcolumn}
\usepackage{graphicx}
\usepackage{epstopdf}
\usepackage{color}
\usepackage{epsf}
\usepackage{epsfig}
\usepackage{graphicx, epic, eepic, color}

\newcommand{\beq}{\begin{equation}}
\newcommand{\eeq}{\end{equation}}

\newcommand{\negminispace}{\kern-.016667em} 

\newcommand{\half}{\kern.083333em}   
\newcommand{\quart}{\kern.0416675em}  
\newcommand{\nhalf}{\kern-.083333em}   
\newcommand{\nquart}{\kern-.0416675em}  

\newcommand{\dual}[1]{{}^\ast\nhalf\nquart#1}

\newcommand\dW{\dual{C}}  

\newcommand\fscalar[1]{{}^\circ{\nquart\nhalf #1}}
\newcommand\fvec[1]{{}^\dagger{\nhalf #1}} 
\newcommand\ftensor[1]{{}^\ddagger{\nquart #1}}

\begin{document}

\title{Spinning Particles in Twisted Gravitational Wave Spacetimes}

\author{Donato \surname{Bini}$^{1,2}$}
\email{donato.bini@gmail.com}
\author{Carmen \surname{Chicone}$^{3,4}$}
\email{chiconec@missouri.edu}
\author{Bahram \surname{Mashhoon}$^{4,5}$}
\email{mashhoonb@missouri.edu}
\author{Kjell \surname{Rosquist}$^{6}$}
\email{kr@fysik.su.se}

\affiliation{
$^1$Istituto per le Applicazioni del Calcolo ``M. Picone'', CNR, I-00185 Rome, Italy\\
$^2$ICRANet, Piazza della Repubblica 10, I-65122 Pescara, Italy\\
$^3$Department of Mathematics, University of Missouri, Columbia, Missouri 65211, USA\\
$^4$Department of Physics and Astronomy, University of Missouri, Columbia, Missouri 65211, USA\\
$^5$School of Astronomy, Institute for Research in Fundamental
Sciences (IPM), P. O. Box 19395-5531, Tehran, Iran\\
$^6$Department of Physics, Stockholm University, SE-106 91 Stockholm, Sweden\\}

\date{\today}

\begin{abstract}
Twisted gravitational waves (TGWs) are nonplanar waves with twisted rays that move along a fixed direction in space. We study further the physical characteristics of a recent class of Ricci-flat solutions of general relativity representing TGWs with wave fronts that have negative Gaussian curvature. In particular, we investigate the influence of TGWs on the polarization of test electromagnetic waves and on the motion of classical spinning test particles in such radiation fields.  To distinguish the polarization effects of twisted waves from plane waves, we examine the theoretical possibility of existence of spin-twist coupling and show that this interaction is generally consistent with our results.
\end{abstract}


\maketitle

\section{Introduction}

In a recent paper~\cite{Bini:2018gbq} about twisted gravitational waves (TGWs), Ricci-flat solutions with diagonal metrics of the form
\begin{equation}\label{S1a}
 ds^2 = - e^{A(u, x)}(dt^2-dz^2) + e^{B(u, x)}\,dx^2 + e^{C(u, x)}\,dy^2\,
\end{equation}
were studied  within the framework of general relativity (GR). Here,  $u := t-z$ is the retarded null coordinate and metric~\eqref{S1a} represents a gravitational wave propagating along the $z$ direction.  The corresponding gravitational field equations were solved in Ref.~\cite{Bini:2018gbq} and it turned out that they contained two separate cases  involving a smooth function $\Psi(u, x)$ that satisfies the partial differential equation 
\beq \label{S1ab}
(\Psi \,\Psi_{uu})_x = 0\,,
\eeq
where $\Psi_u := \partial \Psi/\partial u$, etc.
The first case involved \emph{pp}-waves, namely, plane-fronted waves with parallel rays, with a Petrov type \emph{N} metric of the form
\beq \label{S1b}
ds^2 = -dt^2 + dz^2  + \lambda'^2\,\Psi_{x}^2\, dx^2 + \Psi^2\, dy^2\,,
\eeq
where $\lambda' > 0$ is a constant parameter, while the second case involved nonplanar gravitational waves with a Petrov type \emph{II} metric  such that 
\beq \label{S1}
ds^2 = \Psi^4\,( -dt^2 + dz^2 ) + \lambda^2\,\Psi^4\,\Psi_{x}^2\, dx^2 + \frac{1}{\Psi^2}\, dy^2\,,
\eeq
where $\lambda > 0$ is a constant parameter. 
We remark that these two cases are \emph{not} conformally related to each other. 

The gravitational waves described by Eq.~\eqref{S1} have twisted rays, since the twist tensor 
$\mathbb{T}_{\mu \nu}$ is nonzero in this case,
\beq \label{S2a}
 \mathbb{T}_{\mu \nu} =k_{[\mu ; \nu]} = k_{[\mu , \nu]} \ne 0\,,
\eeq
where $k= \partial_t + \partial_z$ is such that
\beq \label{S2b}
k_\mu\,k^\mu = 0\,, \qquad  k_{\mu ; \nu} + k_{\nu ; \mu} = 0\,, \qquad k_{\mu ; \nu}\,k^{\nu} = 0\,,
\eeq
which therefore  represents a nonexpanding and shearfree null geodesic congruence.  Our TGW solutions belong to Kundt's class~\cite{R1}, see Appendix A.

The twist  tensor for metric~\eqref{S1} has nonzero components $\mathbb{T}_{01} = -\mathbb{T}_{10} = \mathbb{T}_{13} = -\mathbb{T}_{31} = -2\,\Psi^3\,\Psi_x$ and vanishes completely if $\Psi_x = 0$; moreover, 
\beq \label{S2c}
 \mathbb{T}_{\mu \nu}\,\mathbb{T}^{\mu \nu} = 0\,, 
\eeq
in agreement with the fact that the propagation vector $k$ is normal to the wave front, which are $u =$ constant surfaces. That is, a null geodesic congruence is hypersurface-orthogonal if and only if the twist scalar vanishes, which is equivalent to Eq.~\eqref{S2c} as well as $k_{[\mu}\,k_{\nu;\rho]} = 0$. Indeed,  $k_\mu = -\partial u/\partial x^\mu$ for  \emph{pp}-waves, while $k_\mu = \Omega(u, x)\,\partial u/\partial x^\mu$ for twisted gravitational waves. Here, $\Omega_x \ne 0$; otherwise, we have a  \emph{pp}-wave. For metric~\eqref{S1}, $\Omega= - \Psi^4$. Let us note that in general the propagation vector $k$ is fixed up to a nonzero multiplicative constant whose magnitude represents a change in the units of measurement of temporal and spatial intervals corresponding to the freedom in the choice of the affine parameter along the null geodesic world line. 

Consider a Ricci-flat solution of GR such that the corresponding spacetime admits a null Killing vector field $k$ with vanishing twist scalar. Thus $k$ is hypersurface-orthogonal; moreover,  if $k$ is a gradient, then the solution is a \emph{pp}-wave. Assuming that $k$ is \emph{not} a gradient, Dautcourt has shown that the field equations can be solved and reduced to a two-dimensional Poisson equation; furthermore, Dautcourt has studied some of the properties of the resulting gravitational fields~\cite{Daut1, Daut2}. Twisted gravitational waves belong to Dautcourt's class of solutions. 

In the rest of this section, we present the main physical properties of TGWs. In particular, we show that some TGWs have Minkowski spacetime as their background. The Bel-Robinson tensor for TGWs is studied in Section II. The propagation of electromagnetic waves on a TGW background spacetime is treated in Section III. The background has \emph{twisted} null rays; therefore, one might expect that photon spin would couple to the twist and this coupling would cause extra polarization properties of electromagnetic waves as they pass through such a gravitational radiation field. Our results tend to favor photon helicity-twist coupling; hence, to elucidate this interaction further, we study the motion of classical spinning test particles on a certain TGW background in Section IV.  Section V contains a discussion of our results. Throughout, we use units such that $G = c = 1$, unless specified otherwise. 
The signature of the metric is +2 and greek indices run from 0 to 3, while latin indices run from 1 to 3. As in Ref.~\cite{Bini:2018gbq}, all lengths are measured in units of a constant length scale $T_0$ and we set $T_0 = 1$ for the sake of simplicity.  

Let us now concentrate on Eq.~\eqref{S1} that involves TGWs, and note that Eq.~\eqref{S1ab} implies
\beq \label{S2}
\Psi \,\Psi_{uu} = \upsilon(u)\,, 
\eeq
where $\upsilon$ is an arbitrary function of $u$.  For $\upsilon = 0$, we find \emph{simple} Petrov type \emph{D} solutions with $\Psi = u\, f(x) + h(x)$.  
Two such special TGW solutions have been discussed at some length in Ref.~\cite{Bini:2018gbq}. The first special solution is originally due to Harrison~\cite{Harrison:1959zz, DIRC}  and is given by $h(x) = 0$. It can be expressed in the form
\beq \label{S3}
ds^2 = - x^{4/3}\,dt^2 + u^{6/5} dx^2 +  x^{-2/3}\,u^{-2/5} dy^2 + x^{4/3}\,dz^2\,.
\eeq
The second special solution is given by $h/f = x$, where $f$ is a nonzero constant. It can be expressed as
\begin{equation}\label{S4}
 ds^2 = -\lambda_0\,w^4 (dt^2-dz^2) + w^4\, dx^2 + w^{-2} \,dy^2\,,
\end{equation}
where  $\lambda_0 > 0 $ and $w = u + x$.  Comparing Eqs.~\eqref{S1} and~\eqref{S4}, we note that $\Psi$ is in this case $u+x$ up to a proportionality constant. With this $\Psi$, the metric in Eq.~\eqref{S1b} reduces to $(g_{\mu \nu}) = \text{diag}[-1, 1, (u+x)^2, 1]$, which turns out to represent flat spacetime; indeed, this is also true if $u+x$ is replaced by $\chi\,u +\eta\,x$, where $\chi$ and $\eta$ are constant parameters. 

The coordinate systems in these two simple type \emph{D} solutions are admissible over the full range of the spacetime coordinates $x^\mu = (t, \mathbf{x})$, except at curvature singularities which occur when $\bar{w}:= u^{1/5}\,x^{1/3}$ and $w$ vanish in the Harrison and the second solution, respectively.  The coordinate systems are even Lichnerowicz admissible, which means that there are no closed timelike curves in these TGW spacetimes~\cite{Bini:2012ht}. For background material regarding various aspects of the exact solutions of GR representing gravitational radiation, we refer to  Refs.~\cite{R1, Griffiths:2009dfa}.

In connection with the physical interpretation of these solutions, we note that in the general form of our simple type \emph{D} solutions with $\Psi = u\, f(x) + h(x)$, we can set $f = \chi$ and $h(x) = 1 + \eta\,x$, where $\chi$ and $\eta$ are constant parameters and $\eta$ is such that $\lambda^2\,\eta^2 = 1$. We then have a TGW metric of the form
\begin{equation}\label{S5}
 ds^2 = -W^4 (dt^2-dz^2) + W^4\, dx^2 + W^{-2} \,dy^2\,, \qquad W := 1 + \chi\,u + \eta \,x\,,
\end{equation}
which depends on two \emph{arbitrary} constant parameters $\chi$ and $\eta$. It is Ricci-flat and reduces to Minkowski spacetime in the absence of the wave potential $\chi\,u + \eta \,x$, i.e., when
 $\chi = \eta = 0$. It is therefore a TGW on the Minkowski spacetime background. Alternatively, via a simple translation in $x$, say, we could simply replace $x$ by $x-1/\eta$ and thus assume that $W := \chi\,u + \eta \,x$. Then, for $\chi = 0$, our spacetime reduces to  the static Kasner spacetime given by~\cite{Bini:2018gbq}
\begin{equation}\label{S6}
 ds^2 = -x^{4/3} (dt^2-dz^2) + dx^2 + x^{-2/3} \,dy^2\,. 
\end{equation} 
Thus for $\chi \ne 0$, we have a gravitational wave running on this static Kasner background. 
 What could produce such waves on Minkowski and static Kasner backgrounds?
The gravitational radiation emitted by an isolated astrophysical system is expected to have approximately spherical wave fronts very far away from the source.  The spherical waves are locally planar. It is therefore not clear at present what kind of source could generate nonplanar unidirectional gravitational waves. We tentatively assume that TGWs are running cosmological waves. 

It is worthwhile to mention briefly that Harrison's solution~\eqref{S3} can also be written as a TGW on the Minkowski spacetime background, namely,
\beq \label{S7a}
ds^2 = - (1+\eta\,x)^{4/3}\,(dt^2-dz^2) + (1+\chi\,u)^{6/5} dx^2 + (1+\eta\, x)^{-2/3}\,(1+\chi\,u)^{-2/5} dy^2\,,
\eeq
where $\chi$ and $\eta$ are constant parameters. To derive this form of Harrison's metric, we start with Eq.~\eqref{S3} and replace $t$ by $(\eta^{2/3}/\chi)^{5/3}\,t$, $x$ by $(\chi/\eta^{5/3})\,x$, $y$ by $\eta^{-1/3}\,y$ and $z$ by $(\eta^{2/3}/\chi)^{5/3}\,z$; the result is
\beq \label{S7b}
ds^2 = - \frac{1}{\chi^2}\,x^{4/3}\,(dt^2-dz^2) + \frac{1}{\eta^2}\,u^{6/5} dx^2 +  x^{-2/3}\,u^{-2/5} dy^2\,.
\eeq
Finally, if in Eq.~\eqref{S7b} we replace $(t, x, y, z)$ by $(1+\chi\,t, 1+\eta\,x, y, \chi\,z)$, we get Eq.~\eqref{S7a}.

Let us next return to metric~\eqref{S1} and introduce the fundamental static observers that are at rest in space in this general TGW spacetime. They carry the natural tetrad frame $e^{\mu}{}_{\hat \alpha}$ such that 
\begin{eqnarray}\label{S8}
e_{\hat 0} = \frac{1}{\Psi^2}\,\partial_t\,, \qquad
e_{\hat 1} = \frac{1}{\lambda\,\Psi^2\,\Psi_x}\,\partial_x\,,\qquad
e_{\hat 2} =  \Psi\,\partial_y\,, \qquad
e_{\hat 3} = \frac{1}{\Psi^2}\,\, \partial_z\,.
\end{eqnarray}
Here, the 4-velocity of the static observer at the spatial position $(x_0, y_0, z_0)$ is given by $U^\mu = dx^\mu/d\tau$, where $\tau$ is the proper time along the static observer's world line and
\beq \label{S9}
\frac{dt}{d\tau} = \frac{1}{\Psi^2(t - z_0, x_0)}\,.
\eeq
The fundamental static observers are in general accelerated with 4-acceleration $\bar{A}^\mu$ given by
\beq \label{S10}
\bar{A} = \nabla_{e_{\hat 0}}e_{\hat 0}= \frac{2}{\Psi^5}\,\left(\frac{1}{\lambda^2\,\Psi_x}\,\partial_x - \Psi_u\,\partial_z\right)\,.
\eeq
It is straightforward to show that the natural tetrad frame $e^{\mu}{}_{\hat \alpha}$ of the static observers
 is Fermi-Walker transported along these observers' world lines; that is, 
\beq \label{S11}
\frac{De^{\mu}{}_{\hat i}}{d\tau} = (\bar{A}_\alpha\, e^{\alpha}{}_{\hat i})\,e^{\mu}{}_{\hat 0}\,.
\eeq
The twist tensor as measured by the fundamental static observers is 
\beq \label{S12}
\mathbb{T}_{\hat \alpha \hat \beta} = \mathbb{T}_{\mu \nu}\, e^{\mu}{}_{\hat \alpha}\, e^{\nu}{}_{\hat \beta}\,.
\eeq
More explicitly, the nonzero components of $\mathbb{T}_{\hat \alpha \hat \beta}$ are given by
\beq \label{S13}
\mathbb{T}_{\hat 0\hat 1} = -\mathbb{T}_{\hat 1\hat 0} = \mathbb{T}_{\hat 1 \hat 3} = -\mathbb{T}_{\hat 3 \hat 1} = - \frac{2}{\lambda\,\Psi}\,.
\eeq
Thus, in close analogy with the electromagnetic field tensor, one can say that the measured twist tensor has an ``electric" component 
$\mathbb{E}$ in the $ x$ direction and a ``magnetic" component $\mathbb{B}$ in the $y$ direction that have the same magnitude given by $2/(\lambda\, \Psi)$.

There are four algebraically independent scalar polynomial curvature invariants in any Ricci-flat   solution of GR. These can be expressed in terms of the complex quantities
\begin{equation}\label{S14}
I_1 = R_{\mu \nu \rho \sigma}\,R^{\mu \nu \rho \sigma} - i R_{\mu \nu \rho \sigma}\,R^{*\,\mu \nu \rho \sigma}\,
\end{equation}  
and
\begin{equation}\label{S15}
I_2 = R_{\mu \nu \rho \sigma}\,R^{\rho \sigma \alpha \beta}\,R_{\alpha \beta}{}^{\mu \nu} + i R_{\mu \nu \rho \sigma}\,R^{\rho \sigma \alpha \beta}\,R^{*}{}_{\alpha \beta}{}^{\mu \nu}\,.
\end{equation}
For metric~\eqref{S1}, $I_1$ and $I_2$ are real and are given by
\begin{equation}\label{S16}
I_1= \frac{12}{\lambda^4\,\Psi^{12}}\,, \qquad I_2=-\frac{12}{\lambda^6\,\Psi^{18}}\,,
\end{equation}
so that $I_1^3 = 12\,I_2^2$, since the spacetime is algebraically special of Petrov type \emph{II}. It follows from Eq.~\eqref{S16} that the TGW spacetime has a curvature singularity if $\Psi \to 0$. 

Is it possible to have a nonsingular TGW spacetime? If $ \upsilon(u)$ in Eq.~\eqref{S2} is positive and bounded above zero, then $\Psi(u, x)$  does not vanish during its evolution and the corresponding TGW spacetime is free of curvature singularities. To see this, we note that
\begin{equation}\label{S17}
(\Psi^2)_{uu} = 2\,[\Psi_u^2 + \upsilon(u)] > 0\,,
\end{equation}
so that for any given $x$,  $\Psi^2$ as a function of $u$ is  \emph{convex} and could possibly have a zero at its minimum when $ u = u_{\rm min}$. But this would lead to a contradiction with Eq.~\eqref{S2}, since $ \upsilon(u_{\rm min}) > 0$ by assumption. 

An explicit example of this nonsingular situation can be worked out for $\upsilon = k_0$, where $k_0 > 0$ is a constant. Assuming that $\Psi > 0$, Eq.~\eqref{S2} can be integrated once with the result that 
$\Psi_u^2 - 2k_0\,\ln \Psi = 2\,E(x)$, where the total ``energy" $E(x)$ is a finite integration function. It is convenient to write this result as $\Psi_u^2 =  2k_0\,(\hat{E} + \ln \Psi )$, where $\hat{E}(x) = E(x)/k_0$, and interpret  the variation of $\Psi$ as a function of $u$ in terms of a one-dimensional motion of a classical  particle in an effective potential $ -k_0\,\ln \Psi$ with a turning point at $\Psi = \exp(-\hat{E})$. Thus for any given $x$, $-\infty < x < \infty$, $\Psi(u, x)$ monotonically decreases from $\Psi = \infty$, has a minimum at $u_{\rm min}$, where $\Psi(u_{\rm min}) = \exp(-\hat{E}(x))$, and monotonically increases back to $\infty$. Hence, 
\begin{equation}\label{S17}
{\rm erfi}\left(\sqrt{\hat{E}(x)+ \ln{\Psi(u, x)}}\right) = \sqrt{\frac{2k_0}{\pi}}\,(u - u_{\rm min})\,\exp{(\hat{E}(x))}\,,
\end{equation}
where the \emph{imaginary error function} is defined by
\begin{equation}\label{S18}
{\rm erfi}(x) := \frac{2}{\sqrt{\pi}}\,\int_{0}^{x} e^{t^2}\,dt\,. 
\end{equation}

Finally, we note that for any constant $u$, $u = u_0$, say, the spacetime metric $ds^2$ of the TGW~\eqref{S1} reduces to the metric of the wave front $d\sigma^2$,
\begin{equation}\label{S19}
d\sigma^2 = \lambda^2\,\psi^4\,d\psi^2 + \frac{1}{\psi^2}\,dy^2\,, \qquad \psi(x) := \Psi(u_0, x)\,,
\end{equation}
which has negative Gaussian curvature $K_G$,
\begin{equation}\label{S20}
K_G = -\frac{4}{\lambda^2\,\psi^6}\,. 
\end{equation}

\section{Bel-Robinson Tensor for TGWs}

The completely symmetric and traceless Bel-Robinson tensor $T_{\mu \nu \rho \sigma}$ is the natural gravitational analog of the electromagnetic energy-momentum tensor for spacetime domains that are free of sources of gravity. It provides a local measure of the super-energy and super-momentum contained in an arbitrary Ricci-flat region of the gravitational field and is given by 
\beq
\label{BR1}
 T^{\alpha\beta\gamma\delta}
 = \frac{1}{2}
   ( C^{\alpha\mu\gamma\nu}\, C^{\beta}{}_{\mu}{}^{\delta}{}_{\nu}
   + {}^* C^{\alpha\mu\gamma\nu} \,{}^* C^{\beta}{}_{\mu}{}^{\delta}{}_{\nu} )
\,.
\eeq
For a \emph{geodesic} observer with an orthonormal tetrad  in a Ricci-flat gravitational field, the super-energy density and the super-momentum density of the gravitational field as measured by the observer, up to a positive constant scale factor, are given by the appropriate projections of the Bel-Robinson tensor on the observer's tetrad~\cite{Mashhoon:1996wa, Mashhoon:1998tt, bini-jan-min, Clifton:2016mxx} 
\beq \label{BR2}
\mathcal{E} = T^{\hat 0 \hat 0 \hat 0 \hat 0} = \frac{1}{2}\,\text{tr}(E^2 + B^2)\,, \qquad \mathcal{P}^{\hat i} = T^{\hat 0 \hat i \hat 0 \hat 0} = -\epsilon_{\hat i \hat j \hat k}\,(EB)_{\hat j \hat k}\,,
\eeq
respectively. Here,  the symmetric and traceless $3\times 3$ matrices $(E_{\hat i \hat j})$ and $(B_{\hat i \hat j})$  represent the ``electric" and ``magnetic" components of the Weyl curvature tensor, respectively. It is possible to show that $\mathcal{E}$ is positive definite and  the gravitational super-momentum density 4-vector $(\mathcal{E}, \mathcal{P}^{\hat i})$ is always timelike or null. For a more thorough description of $E$ and $B$,  see Appendix B. 

In the rest of this section, we will consider the Bel-Robinson tensor for the two simple type \emph{D}  solutions given by Eqs.~\eqref{S3}  and~\eqref{S4}, respectively.  The timelike geodesic equations are quite complicated even for these simple TGW solutions~\cite{Bini:2018gbq, Bini:2017qnd}; therefore, to have tractable results we resort to the static \emph{accelerated} observers that always remain at rest in space and carry orthonormal tetrads  $e^{\mu}{}_{\hat \alpha}$ discussed in the previous section. 

\subsection{Bel-Robinson tensor for Harrison's solution}

Let us first consider the Harrison metric~\eqref{S3} and
choose as fiducial observers the static observes that carry the orthonormal tetrad 
\beq \label{BR3}
U = e_{\hat 0} = x^{-2/3}\,\partial_t\,, \qquad e_{\hat 1} = u^{-3/5}\, \partial_x\,,\qquad 
e_{\hat 2}= x^{1/3} u^{1/5}\, \partial_y\,,\qquad 
e_{\hat 3}= x^{-2/3}\,\partial_z\,.
\eeq
In this case, the nonzero frame components of the electric and magnetic parts of the Weyl tensor are given by
\begin{eqnarray}\label{BR4}
E^{\hat1 \hat1}&=&-\frac{2}{9\, x^2 u^{6/5}}+\frac{6}{25\, x^{4/3}u^2}\,,\nonumber\\
E^{\hat 2 \hat 2}&=&-\frac{2}{9\, x^2 u^{6/5}}-\frac{6}{25\, x^{4/3}u^2}\,,\nonumber\\
E^{\hat 3 \hat 3}&=& \frac{4}{9 x^2 u^{6/5}}\,,\nonumber\\
E^{\hat1 \hat 3}&=& E^{\hat3 \hat 1} = \frac{2}{5\,  x^{5/3} u^{8/5}}\,
\end{eqnarray}
and
\begin{eqnarray}\label{BR5}
B^{\hat 1 \hat 2}&=&B^{\hat 2 \hat 1} = -\frac{6}{25\, x^{4/3} u^2} \,,\qquad
B^{\hat 2 \hat 3}= B^{\hat 3 \hat 2}= -\frac{2}{5\, x^{5/3} u^{8/5}}\,,
\end{eqnarray}
respectively. We note that these components all diverge at $\bar{w}= u^{1/5}\,x^{1/3} = 0$,  where Harrison's spacetime is singular.
It follows that
\begin{equation}\label{BR6}
   \mathcal{E} = \frac{4\,(625\,u^{8/5} + 1350\, u^{4/5} x^{2/3} + 486\, x^{4/3})}
                      {16875\, u^4 x^4}
\end{equation}
and 
\begin{equation}\label{BR7}
   \mathcal{P}^{\hat i} =\frac{4\,(25\, u^{4/5}+ 18\, x^{2/3})}{1875\, u^4 x^4}
        (- 5\,u^{2/5} x^{1/3}, 0, 3\,x^{2/3})\,. 
\end{equation}

 As is well known, the gravitational Poynting vector depends upon the observer. Indeed, in a vacuum spacetime of type \emph{D}, there exist at each event in spacetime two special (tidal) directions in space such that under boosts in these directions the Poynting vector vanishes for timelike observers and the gravitational super-energy density remains invariant. This circumstance, which has a direct analog in electromagnetism, can be simply illustrated in Kerr spacetime, see Ref.~\cite{Mashhoon:1998tt} and the references cited therein.  

The super-momentum density 4-vector $\mathcal{P}^{\hat \alpha} = (\mathcal{E},\mathcal{P}^{\hat i})$ is timelike, since its squared norm is negative; that is, 
\begin{equation}\label{BR8}
\mathcal{P}^{\hat \alpha} \mathcal{P}_{\hat \alpha} = -\frac{16}3 \biggl[ \frac1{243\, u^{24/5} x^8} 
        +\frac1{225\, u^{28/5} x^{22/3}}
        +\frac1{625\, u^{32/5} x^{20/3}} \biggr] < 0\,. 
\end{equation}
This is indeed the expected result for a type \emph{D} spacetime~\cite{BoSe}.

\subsection{Bel-Robinson tensor for the second simple TGW solution}

Let us next consider our second simple TGW solution given by Eq. \eqref{S4}. As before, we choose  the static observers with adapted tetrad
\beq\label{BR9}
U=e_{\hat 0}= \frac{1}{\lambda_0^{1/2}\,w^2 }\,\partial_t\,, \qquad e_{\hat 1} =\frac{1}{w^2 }\,\partial_x \,,\qquad 
e_{\hat 2}=w\,\partial_y\,,\qquad e_{\hat 3}= \frac{1}{\lambda_0^{1/2}\,w^2}\,\partial_z\,
\eeq
as our reference observers. The nonzero frame components of the electric and magnetic parts of the Weyl tensor are then given by
\begin{eqnarray}\label{BR10}
E^{\hat1 \hat1}&=& \frac{2(3-\lambda_0)}{\lambda_0\,w^6}\,,\nonumber\\
E^{\hat 2 \hat 2}&=& -\frac{2(3+\lambda_0)}{\lambda_0\,w^6}\,,\nonumber\\
E^{\hat 3 \hat 3}&=&  \frac{4}{w^6}\,,\nonumber\\
E^{\hat1 \hat 3}&=& E^{\hat3 \hat 1} = \frac{6} {\lambda_0^{1/2}\,w^6}\,
\end{eqnarray}
and
\begin{eqnarray}\label{BR11}
B^{\hat 1 \hat 2}&=&B^{\hat 2 \hat 1} = -\frac{6}{\lambda_0\, w^6} \,,\qquad
B^{\hat 2 \hat 3}= B^{\hat 3 \hat 2}=  -\frac{6}{\lambda_0^{1/2}\,  w^6}\,,
\end{eqnarray}
respectively. These components all diverge at $w=0$, where spacetime is singular.  
It follows that
\begin{equation}\label{BR12}
   \mathcal{E} = \frac{12\,(\lambda_0^2+6\lambda_0+6)}{\lambda_0^2\, w^{12}}
\end{equation}
and
\begin{equation}\label{BR13}
   \mathcal{P}^{\hat i} = \frac{36\,(\lambda_0+2)}{\lambda_0^2\, w^{12}}
                          (- \lambda_0^{1/2}, 0, 1)\,. 
\end{equation}

The super-momentum density 4-vector is timelike in this case as well, since its squared norm is given by
\begin{equation}\label{BR14}
   \mathcal{P}^{\hat \alpha} \mathcal{P}_{\hat \alpha}
    = -\frac{144\,(\lambda_0^2 + 3\,\lambda_0 +3)}{\lambda_0^2 \,w^{24}} < 0\,. 
\end{equation}

For a plane gravitational wave, the super-momentum density 4-vector is \emph{null} and the gravitational Poynting vector is along the direction of wave propagation~\cite{Mashhoon:1996wa, Mashhoon:1998tt}. Based on the two simple examples presented in this section, we find that for a TGW the super-momentum density 4-vector is timelike and the gravitational Poynting vector has a   
component in the $x$ direction as well, a circumstance that is consistent with the oblique character of the cosmic jet in a TGW spacetime~\cite{Bini:2018gbq, Bini:2017qnd}.

\section{Propagation of Electromagnetic waves in TGW Spacetimes}

The polarization properties of electromagnetic waves change as they propagate through a gravitational radiation field. Previous studies of this prediction of GR have involved plane waves; therefore, it is interesting to investigate the influence of TGWs on the polarization of electromagnetic waves. 
 
On general grounds, we might expect that in a TGW background the photon spin would couple to the twist of the gravitational null rays resulting in the phenomenon of spin-twist coupling. To investigate this possibility, we consider the perturbation of a TGW spacetime via the propagation of a test electromagnetic radiation field. To linear order in the perturbation, the back reaction of the electromagnetic wave on the spacetime may be neglected and Maxwell's equations take the form $F_{[\mu \nu , \rho]} = 0$ and 
$(\sqrt{-g}\,F^{\mu \nu})_{,\nu} = 0$ on a background TGW spacetime.  A convenient method of treating these equations involves replacing the gravitational field by a hypothetical optical medium that occupies Cartesian space~\cite{Sk, Pl, Fe, BM1, BM2, BM3}. Using the natural decompositions $F_{\mu \nu} \to (\mathbf{E}, \mathbf{B})$ and 
$\sqrt{-g}\,F^{\mu \nu} \to (-\mathbf{D}, \mathbf{H})$, one arrives at the standard sourcefree Maxwell's equations in a gyrotropic medium in Euclidean space with Cartesian coordinates and constitutive relations
\begin{equation}\label{P1}
D_i = \epsilon_{ij}\,E_j - (\mathbf{G} \times \mathbf{H})_i\,, \qquad B_i = \mu_{ij}\,H_j + (\mathbf{G} \times \mathbf{E})_i\,,
\end{equation} 
where the conformally invariant properties of the optical medium are given by
\begin{equation}\label{P2}
\epsilon_{ij} = \mu_{ij} = -\sqrt{-g}\,\frac{g^{ij}}{g_{tt}}\,, \qquad G_i = - \frac{g_{ti}}{g_{tt}}\,.
\end{equation} 
For Minkowski spacetime, $g_{\mu \nu} = \eta_{\mu \nu}$, $\epsilon_{ij} =  \mu_{ij} = \delta_{ij}$ and the gyration vector vanishes $(\mathbf{G} = 0)$, as expected. The constitutive relations~\eqref{P1} are derived in Appendix C. 

Let us briefly digress here and consider an inertial frame with coordinates $x^\mu = (t, \mathbf{x})$ in Minkowski spacetime and a plane monochromatic electromagnetic wave of frequency $\omega$  that propagates along the $z$ direction. Taking advantage of the linearity of Maxwell's equations, we can express the corresponding electric and magnetic fields in \emph{complex} form using the circular polarization basis vectors as   
\begin{equation}\label{P3}
\mathbf{e}_{\pm} = a_{\pm}\,(\hat{\mathbf{x}} \pm i\, \hat{\mathbf{y}})\,e^{-i\omega\,(t-z)}\,, \qquad
\mathbf{b}_{\pm} = \mp\, i\, a_{\pm}\,(\hat{\mathbf{x}} \pm i\, \hat{\mathbf{y}})\,e^{-i\omega\,(t-z)}\,,
\end{equation} 
where $a_{+}$ and $a_{-}$ are constant complex amplitudes for positive and negative helicity radiation, respectively, and $\hat{\mathbf{x}}$ indicates a unit vector in the $x$ direction, etc. When projected onto the orthonormal tetrad frame of an observer, the \emph{real} parts of our complex fields would correspond to measurable quantities. It is important to note that for \emph{positive-helicity} radiation,  
\begin{equation}\label{P4}
\mathbf{e}_{+} + i\, \mathbf{b}_{+} = 2\,a_{+}\,(\hat{\mathbf{x}} + i\, \hat{\mathbf{y}})\,e^{-i\omega\,(t-z)}\,, \qquad
\mathbf{e}_{+} - i\, \mathbf{b}_{+}  = 0\,,
\end{equation} 
while for \emph{negative-helicity} radiation, 
\begin{equation}\label{P5}
\mathbf{e}_{-} + i\, \mathbf{b}_{-}  = 0\,, \qquad \mathbf{e}_{-} - i\, \mathbf{b}_{-} = 2\,a_{-}\,(\hat{\mathbf{x}} - i\, \hat{\mathbf{y}})\,e^{-i\omega\,(t-z)}\,.
\end{equation} 
Thus in terms of complex fields, $\mathbf{e} + i\, \mathbf{b}$ essentially represents an electromagnetic wave with positive helicity, while $\mathbf{e} - i\, \mathbf{b}$ essentially represents a wave with negative helicity. It is clear that one will arrive at this same conclusion if one starts from a linear polarization basis for the complex fields instead of the circular polarization basis employed in Eq.~\eqref{P3}. 

Returning now to the optical medium in Cartesian space that supplants the gravitational field, we introduce the Riemann-Silberstein vectors
\begin{equation}\label{P6}
\mathbf{F}^{\pm} = \mathbf{E} \pm i\,\mathbf{H}\,, \qquad  \mathbf{S}^{\pm} = \mathbf{D} \pm i\,\mathbf{B}
\end{equation} 
in terms of complex fields, so that Maxwell's equations can now be written as 
\begin{equation}\label{P7}
\nabla \times \mathbf{F}^{\pm} = \pm\, i\,\frac{\partial \mathbf{S}^{\pm}}{\partial t}\,, \qquad \nabla \cdot \mathbf{S}^{\pm} = 0\,, 
\end{equation} 
where 
\begin{equation}\label{P8}
S^{\pm}_p = \epsilon_{pq}\,F^{\pm}_q  \pm i\, (\mathbf{G} \times \mathbf{F^{\pm}})_p\,. 
\end{equation} 
These field equations completely decouple for different helicity states in this linear perturbation treatment.  Moreover, the equation for  temporal evolution implies $\partial_t(\nabla \cdot \mathbf{S}^{\pm}) = 0$, which means that once the second equation of display~\eqref{P7} is satisfied at any given time, then it is valid for all time. 

In Minkowski spacetime, $\mathbf{F}^{+}$ is the amplitude for radiation of positive helicity, while $\mathbf{F}^{-}$ represents radiation of negative helicity, as explained before. Similarly, in a scattering situation involving asymptotically flat spacetimes, Eqs.~\eqref{P7}--\eqref{P8} imply that the helicity of the radiation is preserved in gravitational scattering. More generally, $\mathbf{F}^{+}$ and $\mathbf{F}^{-}$ may be considered helicity amplitudes in an arbitrary gravitational medium as a natural extension of the standard notion for fields in Minkowski spacetime. This is the approach we adopt in the rest of this section for the solution of Maxwell's equations in curved spacetime. 

Let us first consider the propagation of test electromagnetic waves through a weak gravitational radiation field with metric $g_{\mu \nu } = \eta_{\mu \nu} + h_{\mu \nu}$. Here, $h_{\mu \nu}$ represents a gravitational wave of frequency $\Omega_g$ and wave vector $\mathbf{K}_g$ according to the static inertial observers in spacetime. Let $\hat{h}$, $0<\hat{h}\ll1$ denote the order of magnitude of the absolute value of the nonzero elements of  $h_{\mu \nu}$. In the TT gauge for gravitational radiation, the gravitational field is equivalent to an optical medium with $\epsilon_{ij} = \delta_{ij} -h_{ij}$ and $\mathbf{G} = 0$. A monochromatic electromagnetic wave of frequency $\omega$ and wave vector $\mathbf{k}$ propagating through this weak gravitational wave would pick up two components, each of relative magnitude $\sim \hat{h}$, with frequencies $\omega \pm \Omega_g$ and corresponding wave vectors $\mathbf{k}\pm \mathbf{K}_g$, see Appendix A of Ref.~\cite{MaGr}. Naturally, the polarization of the electromagnetic wave would be affected by terms of order $\hat{h}$. Indeed, a more detailed analysis of this problem has been carried out in the eikonal approximation, where the general relativistic rotation of the plane of polarization of electromagnetic waves (Skrotskii effect~\cite{Sk}) has been calculated along null geodesics in spacetimes containing weak gravitational waves emitted by localized astrophysical sources~\cite{Kopeikin:2001dz}. Extending the analysis to exact plane waves, we would expect that the polarization of test electromagnetic waves would change accordingly as they propagate through plane-wave spacetimes. 

Next, we turn to exact twisted gravitational wave solutions. To simplify matters, we choose a variant of  the second special TGW solution given by $\Psi = u\,f(x) + h(x)$ with $f = \chi$ and $h = \eta\,x$, where $\chi$ and $\eta$ are constants. Equation~\eqref{S1} for the general TGW spacetime interval now takes the form
\begin{equation}\label{P8a}
ds^2 = w^4\,(-dt^2+dz^2) +\lambda^2\,\eta^2\, w^4\,dx^2+ w^{-2}\,dy^2\,, 
\end{equation} 
where $w := \chi\,u + \eta \,x$. Let us choose constants such that  $\lambda^2\,\eta^2 = 1$; then,
the spacetime interval that we employ in this section expressed in Cartesian coordinates is represented by
\begin{equation}\label{P9}
(g_{\mu \nu}) = \text{diag}(-w^4, w^4, w^{-2}, w^4)\,, \qquad w =  \chi\,u + \eta \,x\,,
\end{equation} 
where $w = 0$ indicates the curvature singularity of this TGW spacetime. We note that for $\eta = 0$ we have a plane wave, whereas for $\eta \ne 0$ we have a TGW.  It follows that the corresponding optical medium is given by
\begin{equation}\label{P10}
(\epsilon_{ij}) = \text{diag}(w^{-3}, w^3, w^{-3})\,, \qquad \mathbf{G} = 0\,.
\end{equation} 
The optical properties of this medium are independent of the $y$ coordinate; therefore, to solve Maxwell's equations~\eqref{P7} in this case we assume
\begin{equation}\label{P11}
\mathbf{F}^{\pm}(t, \mathbf{x}) = e^{i\,K_{y}^{\pm}\,y}\, \mathbf{f}^{\pm}(w)\,,
\end{equation} 
where $K_{y}^{\pm}$ are \emph{complex} constants. Moreover, we set $\chi = 1$ for the sake of simplicity and assume $\eta \ne 0$.	Then, it follows from the equation for temporal evolution in display~\eqref{P7} that
\begin{equation}\label{P12}
i\,K_{y}^{\pm}\,f_{z}^{\pm} + \frac{df_{y}^{\pm}}{dw} = \pm\,i\, \frac{d(w^{-3}\,f_{x}^{\pm})}{dw}\,,
\end{equation} 
\begin{equation}\label{P13}
- \frac{d(f_{x}^{\pm} +\eta\, f_{z}^{\pm})}{dw} = \pm\,i\, \frac{d(w^{3}\,f_{y}^{\pm})}{dw}\,,
\end{equation}
\begin{equation}\label{P14}
\eta\, \frac{df_{y}^{\pm}}{dw} - i\,K_{y}^{\pm}\,f_{x}^{\pm}  = \pm\,i\, \frac{d(w^{-3}\,f_{z}^{\pm})}{dw}\,.
\end{equation}
Furthermore, the temporal ``boundary condition" $\nabla \cdot \mathbf{S}^{\pm} = 0$ implies
\begin{equation}\label{P15}
\eta\,\frac{d(w^{-3}\,f_{x}^{\pm})}{dw} + i\,K_{y}^{\pm}\,w^3\,f_{y}^{\pm} -  \frac{d(w^{-3}\,f_{z}^{\pm})}{dw} = 0\,.
\end{equation}

To find the solution of Eqs.~\eqref{P12}--\eqref{P15}, let us first assume that $K_{y}^{\pm} = 0$. Then, the electromagnetic field depends only upon $w$ and it is straightforward to show that 
\begin{equation}\label{P16}
F_{x}^{\pm} = c_{x}^{\pm}\,w^3 + \frac{1}{\eta^2}\,c_{y}^{\pm}\,,
\end{equation}
\begin{equation}\label{P17}
F_{y}^{\pm} = \pm\,i\,\left(c_{x}^{\pm} + \frac{1}{\eta^2}\, c_{y}^{\pm}\,w^{-3} + \eta\, c_{z}^{\pm}\right)\,,
\end{equation}
\begin{equation}\label{P18}
F_{z}^{\pm} = \frac{1}{\eta}\,c_{y}^{\pm} + c_{z}^{\pm}\,w^3  \,,
\end{equation}
where $c_{x}^{\pm}$, $c_{y}^{\pm}$ and $c_{z}^{\pm}$ are integration constants. To have a finite \emph{perturbation} at $w = 0$, we must assume that $c_{y}^{\pm} = 0$, in which case $F_{x}^{\pm}$ and $F_{z}^{\pm}$ are proportional to $w^3$ and $F_{y}^{\pm}$ is simply a constant. 

Let us next assume that complex constants $K_{y}^{\pm}$ are such that $K_{y}^{\pm} \ne 0$. Then, Eq.~\eqref{P13} can be simply integrated once, but the corresponding integration constant must be set equal to zero for the sake of consistency of Eqs.~\eqref{P12}--\eqref{P15}. The resulting expression for $f_{y}^{\pm}$ can be substituted in the other equations; in this way, it is straightforward to see that the general solution for $K_{y}^{\pm} \ne 0$ takes the form
\begin{equation}\label{P19}
f_{x}^{\pm} = -\left[\frac{1}{2}\,(\eta-\frac{1}{\eta})\,\alpha^{\pm}\,e^{-\mathcal{U}^{\pm}} - \beta^{\pm}\,e^{\mathcal{U}^{\pm}}\right]\,w^3\,,
\end{equation}
\begin{equation}\label{P20}
f_{y}^{\pm} = \pm\,i\,\left[\frac{1}{2}\,(\eta+\frac{1}{\eta})\,\alpha^{\pm}\,e^{-\mathcal{U}^{\pm}} + \beta^{\pm}\,e^{\mathcal{U}^{\pm}}\right]\,,
\end{equation}
\begin{equation}\label{P21}
f_{z}^{\pm} = \alpha^{\pm}\,e^{-\mathcal{U}^{\pm}}\,w^3\,,
\end{equation}
where $\alpha^{\pm}$ and $\beta^{\pm}$ are constants of integration and $\mathcal{U}^{\pm}$ is given by
\begin{equation}\label{P22}
\mathcal{U}^{\pm} = \pm\,\frac{1}{4\,\eta}\,K_{y}^{\pm}\,w^4\,.
\end{equation}
For fixed $x$ and $z$, $w$ is essentially the temporal parameter and with complex constants 
$\alpha^{+} = \alpha^{-}$, $\beta^{+} = \beta^{-}$ and $K_{y}^{+} = K_{y}^{-}$, it is clear that 
$\mathbf{F}^{+}$ is rather different from $\mathbf{F}^{-}$, which represent the helicity-dependent propagation of an electromagnetic wave along the $y$ direction. 

The particular solution of Maxwell's equations presented here does not have a proper limit for $\eta = 0$. On the other hand, our simple TGW solution~\eqref{P8a} reduces to a plane 
gravitational wave for $\eta = 0$~\cite{Bini:2018gbq}. It is useful to study electromagnetic wave motion in this limit as well. 

For $\eta = 0$ and $\chi = 1$, the metric in Eq.~\eqref{P8a} reduces to 
\begin{equation}\label{P23}
ds^2 = - u^4\,(dt^2- dz^2) + u^4\,dx^2 + u^{-2}\,dy^2\,.
\end{equation}
This linearly polarized plane gravitational wave spacetime can be replaced by an optical medium, namely, 
\begin{equation}\label{P24}
(\epsilon_{ij}) = \text{diag}(u^{-3}, u^3, u^{-3})\,, \qquad \mathbf{G} = 0\,,
\end{equation} 
where $u := t-z$, as before, and $u = 0$ denotes the curvature singularity of this spacetime.  Using the same approach as above, a detailed analysis of the propagation of electromagnetic radiation in this plane gravitational wave background reveals that for $K_{y}^{\pm} = 0$, we have the highly degenerate solution
\begin{equation}\label{P25}
F_{x}^{\pm}(u) \pm \,i\,u^{3}\,F_{y}^{\pm}(u) = 0\,, \qquad F_{z}^{\pm}(u) = \tilde{c}_{z}^{\pm}\,u^3\,,
\end{equation}
where $\tilde{c}_{z}^{\pm}$ are constants. For $K_{y}^{\pm} \ne 0$, however, the only possible exact solution of the electromagnetic perturbation equations of the form~\eqref{P11} in this case is given by 
$\mathbf{F}^{\pm} = 0$. Thus an analog of solution~\eqref{P19}--\eqref{P21} does not exist in this case. 

How are TGWs different from plane or \emph{pp}-waves in terms of their influence on the variation of polarization of electromagnetic waves? To address this basic issue, let us consider the propagation of massless test fields with spin on a TGW background. In view of the existence of spin-rotation coupling~\cite{MK, DSH, SRC}, it is natural to expect that the spin of the test field  would couple to the twist of the gravitational radiation background. In close analogy with the spin-rotation coupling, if there is a spin-twist coupling according to the fiducial static observers, we expect it to be \emph{proportional} to $S^{\mu \nu}\, \mathbb{T}_{\mu \nu}$, where 
$S^{\mu \nu}$ is the spin tensor of the massless field such that $S^{\mu \nu}\,e_{\nu\, \hat 0} = 0$. It is possible to define a corresponding spin vector $S^\mu$ via
\beq \label{T1}
S^\mu = - \frac{1}{2}\,e^{\mu \nu \rho \sigma}\,e_{\nu\, \hat 0}\, S_{\rho \sigma}\,, \qquad S_{\mu \nu} = e_{\mu \nu \rho \sigma}\,e^{\rho}{}_{\hat 0}\,S^\sigma\,,
\eeq
where $e_{\mu \nu \rho \sigma}$ denotes the alternating tensor given by
$e_{\mu \nu \rho \sigma} = \sqrt{-g}\,\epsilon_{\mu \nu \rho \sigma}$. Here, $\epsilon_{\mu \nu \rho \sigma}$ is the  totally antisymmetric symbol with $\epsilon_{0123} = 1$. 
Thus the spin-twist coupling would be of the form
\beq \label{T2}
S^{\mu \nu}\, \mathbb{T}_{\mu \nu} = - e^{\alpha \beta \gamma \delta}\,e_{\beta \,\hat 0}\,S_\alpha\,\mathbb{T}_{\gamma \delta} = - S_\alpha \mathbb{V}^\alpha\,,
\eeq
where we have defined a \emph{twist vector} 
$\mathbb{V}^\mu$  for TGW spacetimes along the world lines of the static observers such that
\beq \label{T3} 
\mathbb{V}^\mu := e^{\mu \nu \rho \sigma}\,e_{\nu\, \hat 0}\, \mathbb{T}_{\rho \sigma}\,.
\eeq
 For our TGW with metric~\eqref{P9}, we find that $\mathbb{V}^\mu$ is given by $\mathbb{V} = -4\,\eta\,\partial_y$. According to the static fiducial observers, spin-twist coupling would involve an interaction proportional to $-S_{\hat \mu} \,\mathbb{V}^{\hat \mu} = 4\,\eta\,S^{\hat 2}/w$, so that the spin of the particle in the $y$ direction is involved.
Does such an interaction indeed occur in the motion of spinning test particles in TGW spacetimes? 

Imagine the propagation of massless test fields on a background TGW spacetime. In the eikonal limit, we expect that the propagation of such a test field with spin would correspond to the motion of a massless spinning test particle, which can be consistently described in GR using the Mathisson-Papapetrou equations only when the Frenkel-Pirani supplementary condition is imposed~\cite{Mash}.  The particle then follows a null geodesic and its spin is along its path with two possible helicity states such that helicity is conserved in an orientable spacetime~\cite{Mash}.  The wave amplitude then propagates along this null geodesic and the eikonal approximation scheme describes the variation of the linear polarization state of the wave as the wave amplitude changes along the null geodesic world line.   Thus in this approximation, the spin of the particle generally couples to the twist in a TGW spacetime if the particle moves along the $y$ direction. That is, except for motion along certain special null geodesics that are spatially confined to the $(x, z)$ surface and do not move in the $y$ direction, we should expect the presence of spin-twist coupling. Otherwise, the spin vector of the particle is in the $(x, z)$ plane either along or opposite to its motion and thus has no $y$ component to couple to the twist vector. We note that a similar situation occurs in the propagation of test electromagnetic waves in the simple TGW spacetime examined in this section, see Eqs.~\eqref{P16}--\eqref{P22}. In this case, the helicity-twist coupling would be proportional to $S^{\hat 2}$, where $S^{\hat 2}$ would in turn be proportional to $\pm\,K_{y}^{\pm}$; for instance, $\mathcal{U}^{\pm} = \pm\,K_{y}^{\pm}\,w^4/(4\,\eta)$ in Eq.~\eqref{P22} could represent photon helicity-twist coupling. Our analytic solution of Maxwell's equations is rather special; nevertheless, the result is generally consistent with expectations based upon the photon helicity-twist coupling. 

To study further the phenomenon of spin-twist coupling,  the next section is devoted to the classical motion of spinning test particles in the field of a simple weak TGW.

\section{Classical Spinning Test Particles in a TGW background}

To describe the motion of classical spinning test masses in a gravitational field, we must use the Mathisson-Papapetrou-Dixon (MPD) equations at the pole-dipole order. For a detailed account of these equations, see Refs.~\cite{CMP, MS} and the references cited therein. To linear order in spin, the MPD equations for a particle of mass $m$, 4-velocity vector $V^\mu:=dx^\mu/d\tau$, momentum $P^\mu$ and spin tensor $S^{\mu \nu}$ simplify in accordance with an approximation scheme developed in Ref.~\cite{CMP}; that is, the momentum of the spinning particle becomes proportional to its 4-velocity, $P^\mu \approx m\, V^\mu$, when terms of second order in spin are neglected. The equations of motion are then
\begin{equation}\label{M1}
\frac{DV^\mu}{d\tau} \approx -\frac{1}{2\,m}\, R^{\mu}{}_{\nu \alpha \beta}\,V^\nu\,S^{\alpha \beta}\, 
\end{equation}
and
\begin{equation}\label{M2}
\frac{DS^{\mu \nu}}{d\tau} \approx 0\,, 
\end{equation}
with the supplementary condition
\begin{equation}\label{M3}
S_{\mu \nu}\,V^\nu \approx 0\,. 
\end{equation}

It proves useful to introduce the spin vector of the test particle $S^\mu$, $S_\mu\,V^\mu = 0$,  such that
\begin{equation}\label{M4}
S_{\mu \nu} = e_{\mu \nu \rho \sigma}\,V^\rho\,S^\sigma\,, \qquad  S_{\mu} = -\frac{1}{2}\,e_{\mu \nu \rho \sigma}\,V^\nu\,S^{\rho \sigma}\,, 
\end{equation}
where $e_{\mu \nu \rho \sigma}$ is the alternating tensor given by
$e_{\mu \nu \rho \sigma} = \sqrt{-g}\,\epsilon_{\mu \nu \rho \sigma}$, as before.  

Let us define the dual curvature tensor $^{*}R_{\mu \nu \rho \sigma}$ via
\begin{equation}\label{M5}
^{*}R_{\mu \nu \rho \sigma} = \frac{1}{2}\,e_{\mu \nu \alpha \beta}\,R^{\alpha \beta}{}_{\rho \sigma}\,. 
\end{equation}
Then, the main equations of motion are
\begin{equation}\label{M6}
\frac{DV^\mu}{d\tau} \approx \frac{1}{m}\, ^{*}R^{\mu}{}_{\nu\rho \sigma}\,V^\nu\,S^{\rho}\,V^\sigma\,
\end{equation}
and
\begin{equation}\label{M7}
\frac{DS^{\mu}}{d\tau} \approx 0\,. 
\end{equation}
It follows from these equations that $V_\mu\,S^\mu \approx 0$ throughout the motion and the magnitude of spin is conserved, since $S_\mu\,S^\mu$ is a constant of the motion.
Moreover,  in this approximation $S^{\mu}$ undergoes parallel propagation along the test particle's world line.

Dynamics of classical spinning test particles have been studied in the field of weak gravitational plane waves by a number of authors, see Refs.~\cite{Bini:2017dny, Collas:2018wcc} and the references cited therein. We discuss this problem below for twisted waves and compare the result with the case of plane waves. 

Let us consider a TGW metric of the form~\eqref{S5}, namely, 
\beq
\label{M8}
ds^2 =-  W^4 (dt^2-dz^2)+ W^4 dx^2+W^{-2}dy^2
\eeq 
with
\beq \label{M9}
W = 1 + \chi \,u +\eta \,x\,,
\eeq
and the  family of static observers with their natural adapted frame:
\beq \label{M10}
e_{\hat 0}= \frac{1}{W^2}\partial_t\,,\qquad e_{\hat 1}=  \frac{1}{W^2}\partial_x\,,\qquad e_{\hat 2}=  W\partial_y\,,\qquad e_{\hat 3}=  \frac{1}{W^2}\partial_z\,.
\eeq
 Metric~\eqref{M8} represents a plane wave for $\eta = 0$ and a twisted wave for $\eta \ne 0$. We will assume throughout this section that the potential $\phi :=\chi \,u +\eta \, x$ is a small perturbation on Minkowski spacetime background, see Appendix D. That is, $g_{\mu \nu} = \eta_{\mu \nu} + h_{\mu \nu}$, where 
$(h_{\mu \nu})=\text{diag}(-4\,\phi, 4\,\phi, -2\,\phi, 4\,\phi)$, $0 < |\phi| \ll 1$ and the corresponding Christoffel symbols at this linear level of approximation are all constants proportional to the small parameters $\chi$, $0 <|\chi| \ll 1$, and $\eta$,  $0 <|\eta| \ll 1$. It follows that the curvature tensor vanishes at this level of approximation. Thus the test particle is in an accelerated coordinate system in Minkowski spacetime and follows a geodesic of the background TGW in this linear approximation scheme.  The coordinate transformation from the accelerated system to the corresponding inertial system has been worked out in Appendix D. 

It is straightforward to solve the geodesic equation in this case and the result for $V^\mu$ is that, at this level of approximation, it depends linearly upon proper time; that is, 
\begin{eqnarray} \label{M11}
V^0 &\approx& \gamma + 2\,(\chi\, \mathcal{A} - 2 \gamma^2\,\mathcal{B})\,\tau\,,\nonumber\\
V^1 &\approx& \gamma\,v_x - 2\,(\eta\, \mathcal{A} +2 \gamma^2\,v_x\,\mathcal{B})\,\tau\,,\nonumber\\
V^2 &\approx& \gamma\,v_y  +2 \gamma^2\,v_y\,\mathcal{B}\,\tau\,,\nonumber\\
V^3 &\approx& \gamma\,v_z + 2\,(\chi\, \mathcal{A} - 2 \gamma^2\,v_z\,\mathcal{B})\,\tau\,,
\end{eqnarray}
where $\gamma = 1/\sqrt{1-v^2}$, $v^2 = v_x^2 + v_y^2+v_z^2$ and
\beq \label{M12}
\mathcal{A} = 1+\frac{3}{2}\gamma^2\,v_y^2\,, \qquad \mathcal{B} = \chi\,(1-v_z) +\eta\, v_x\,,
\eeq
are simply constants. In the absence of acceleration, i.e. with $\chi = \eta = 0$, the test particle is in an inertial frame with Cartesian coordinates in Minkowski spacetime and has 4-velocity $V^\mu = \gamma (1, v_x, v_y, v_z)$ and spin vector $S^\mu = (\sigma^0, \sigma^x, \sigma^y, \sigma^z)$. One can now simply integrate these equations to find the geodesic path $x^\mu (\tau)$ with the initial conditions that at $\tau = 0$, $x^\mu = (t, x, y, z) = (0,0,0,0)$. Thus the path of the spinning particle is given by
\begin{eqnarray}\label{M13}
t(\tau) &\approx& \gamma\tau + (\chi {\mathcal A}-2 \gamma^2 {\mathcal B})\,\tau^2\,,\nonumber\\
x(\tau) &\approx& \gamma v_x\tau -(\eta {\mathcal A}+2\gamma^2v_x {\mathcal B})\,\tau^2\,,\nonumber\\
y(\tau) &\approx& \gamma v_y\tau+\gamma^2 v_y {\mathcal B}\,\tau^2\,,\nonumber\\
z(\tau) &\approx& \gamma v_z\tau + (\chi {\mathcal A}-2\gamma^2 v_z {\mathcal B})\,\tau^2\,.
\end{eqnarray}
It then follows that along the path of the particle the potential $\phi = \chi\,u + \eta\, x$ is given by
$\phi = \gamma \mathcal{B}\tau$.

In a similar way, one can integrate the equations for the spin vector and one finds
\begin{eqnarray} \label{M14}
S^0 &\approx& \sigma^0 - 2\,\gamma\,\chi\,[(\sigma^0-\sigma^z)(1-v_z) +\sigma^x\,v_x -\frac{1}{2}\sigma^y\,v_y]\,\tau  - 2\,\gamma\,\eta\,(\sigma^0\,v_x+\sigma^x)\,\tau\,,\nonumber\\
S^x &\approx& \sigma^x - 2\,\gamma\,\chi\,[(\sigma^0-\sigma^z)\,v_x +\sigma^x\,(1-v_z)]\,\tau  - 2\,\gamma\,\eta\,(\sigma^0 + \sigma^x\,v_x +\frac{1}{2}\sigma^y\,v_y-\sigma^z\,v_z)\,\tau\,,\nonumber\\
S^y &\approx& \sigma^y + \gamma\,\chi\,[(\sigma^0-\sigma^z)\,v_y +\sigma^y\,(1-v_z)]\,\tau  +\gamma\,\eta\,(\sigma^x\,v_y  + \sigma^y\,v_x)\,\tau\,,\nonumber\\
S^z &\approx& \sigma^z + 2\,\gamma\,\chi\,[(\sigma^0-\sigma^z)(1-v_z) -\sigma^x\,v_x +\frac{1}{2}\sigma^y\,v_y]\,\tau  - 2\,\gamma\,\eta\,(\sigma^x\,v_z+\sigma^z\,v_x)\,\tau\,,
\end{eqnarray}
where $\sigma^0$, $\sigma^x$, etc., are constants such that $\sigma^0 = \boldsymbol{\sigma} \cdot \mathbf{v}$.  It is possible to obtain these same $S^\mu$ components via the spin vector transformation law between the inertial and accelerated systems employing the explicit coordinate transformation derived at the end of Appendix D. Using these results, it is possible to check that the spin vector is orthogonal to $V^\mu$ and $S^\mu S_\mu = \eta_{\alpha \beta}\sigma^\alpha \sigma^\beta$ at the order of approximation under consideration in this section. 

To find measurable quantities, the next step involves the calculation of $V^\mu e_{\mu \hat \alpha} = V_{\hat \alpha} = \Gamma(- 1 ,  \mathcal{V}_x,   \mathcal{V}_y,   \mathcal{V}_z)$ and $S^\mu \, e_{\mu \hat \alpha} = S_{\hat \alpha}$ and the study of their properties. 
In this way, we find
\begin{eqnarray}\label{M15}
\Gamma &\approx& \gamma +2\, (\chi {\mathcal A}-\gamma^2 {\mathcal B})\,\tau\,, \nonumber\\
\mathcal{V}_x &\approx& v_x-\frac{2}{\gamma}\, {\mathcal A}\,(\chi v_x+\eta)\,\tau\,,  \nonumber\\
\mathcal{V}_y &\approx& v_y-\frac{v_y}{\gamma}\,(2\chi{\mathcal A}-3\gamma^2{\mathcal B})\,\tau\,, \nonumber\\
\mathcal{V}_z &\approx& v_z+\frac{2}{\gamma}\,\chi {\mathcal A}(1-v_z)\,\tau\,. 
\end{eqnarray}
Moreover, the frame components of the spin vector can be expressed as
\begin{eqnarray}\label{M16}
S^{\hat t}&\approx&  \sigma^0 + 2\,\gamma\,\chi \left[\sigma^z(1-v_z)-\sigma^x v_x+\frac{1}{2}\,\sigma^y v_y\right]\,\tau- 2\,\gamma\,\eta\,\sigma^x\,\tau\,,\nonumber\\
S^{\hat x} &\approx& \sigma^x-2\,\gamma\,\chi\,(\sigma^0 - \sigma^z)\,v_x\,\tau- 2\,\gamma\,\eta\,\left(\sigma^0+\frac{1}{2}\,\sigma^y v_y-\sigma^z v_z\right)\,\tau\,,\nonumber\\
S^{\hat y} &\approx& \sigma^y + \gamma\,\chi\,(\sigma^0-\sigma^z)v_y\,\tau+\gamma\,\eta\,\sigma^x v_y\,\tau\,,\nonumber\\
S^{\hat z}&\approx& \sigma^z + 2\,\gamma\,\chi\,\left[\sigma^0 (1-v_z)-\sigma^x v_x+\frac{1}{2}\,\sigma^y v_y\right]\,\tau -2\,\gamma\,\eta\,\sigma^x v_z\,\tau\,.
\end{eqnarray}
In these results, the special character of motion in the $y$ direction should be noted: If there is initially no motion in the $y$ direction, i.e. $v_y = 0$, then for $\tau \ne 0$ we find $\mathcal{V}_y \approx 0$ and $S^{\hat y} \approx \sigma^y$.

In the original TGW solution given by Eqs.~\eqref{M8} and~\eqref{M9} with $\chi \ne 0$, we note that the corresponding spacetime represents a plane-fronted gravitational wave if $\eta = 0$ and a nonplanar wave if $\eta \ne 0$; therefore, terms proportional to $\eta$ in Eq.~\eqref{M16} are due to the twist of the gravitational wave. Our results are clearly quite different for twisted waves as compared to plane waves.  Moreover, the spin-twist coupling~\eqref{T2} in the case under consideration takes the form $-S_\mu\,\mathbb{V}^\mu \approx 4\,\eta\,\sigma^y$, which is reflected in parts of Eq.~\eqref{M16}.  

It follows immediately from Eq.~\eqref{M16} that if initially (i.e., at $\tau = 0$), the spin is in the direction of wave propagation $z$ according to the fiducial observers, $\eta$ simply drops out of Eq.~\eqref{M16}. On the other hand, if the spin is initially in the $y$ direction, then in time it rotates along the direction of wave propagation due to the presence of $\eta \ne 0$ if the spinning particle has an initial component of motion in the $y$ direction. Similarly, if the spin is initially in the $x$ direction, then in time it develops components along $y$ and $z$ directions due to the presence of $\eta \ne 0$, provided the spinning test particle has initial components of motion along the $y$ and $z$ directions, respectively. 
 
\section{Discussion}

The main physical properties of TGWs, first introduced in Ref.~\cite{Bini:2018gbq}, have been studied in the first part of the present work. Specifically, we have computed the Bel-Robinson tensor for \emph{simple} TGWs. Furthermore, we have  shown that simple TGWs  can have Minkowski spacetime as their backgrounds. The second part of this paper has been devoted to the influence of simple TGWs on the spins of test particles. In particular,  we have studied the polarization of test electromagnetic waves propagating on such backgrounds.  We have also examined the motion of spinning test particles in the field of a simple weak TGW. The difference between the results for twisted waves in comparison with plane waves can be attributed in part to the coupling of the spin of the test particle with the twist of the background.

\appendix

\section{Metric of TGWs in Kundt's Form}

The purpose of this appendix is to put the general metric~\eqref{S1} of TGWs of Petrov type \emph{II} in the standard Kundt form, see Eq.  (31.6) of Chapter 31 of Ref.~\cite{R1}. Let us first introduce the advanced null coordinate $v' = t+z$ in Eq.~\eqref{S1}, namely,
\beq \label{A1}
ds^2 =- \Psi^4\,du\,dv' + \lambda^2\,\Psi^4\,\Psi_{x}^2\, dx^2 + \frac{1}{\Psi^2}\, dy^2\,, \qquad v' = t + z\,,
\eeq 
where $\Psi(u, x)$ is a general solution of the partial differential Eq.~\eqref{S1ab}. Next, the coordinate transformation 
\beq \label{A2}
(u, v', x, y) \mapsto (u, v, x, y)\,, \qquad  v = \frac{1}{2}\,\Psi^4(u, x)\,v'\,,
\eeq 
turns metric~\eqref{A1} into 
\beq \label{A3}
ds^2 =- 2\,du\,dv + \frac{8v}{\Psi}\,\Psi_u\,du^2 + \frac{8v}{\Psi}\,\Psi_x\,du\,dx + \lambda^2\,\Psi^4\,\Psi_{x}^2\, dx^2 + \frac{1}{\Psi^2}\, dy^2\,.
\eeq
We would now like to write this metric in terms of coordinates $(u, v, X, Y)$ such that it takes the form 
\beq \label{A4}
ds^2 =- 2\,du\,dv - 2\, H\, du^2 -2\,\mathcal{W}_1\,du\,dX + 2\,\mathcal{W}_2\,du\,dY +P^{-2} (dX^2+dY^2)\,, 
\eeq
where $P$ is independent of $v$, $P_v = 0$. Introducing complex quantities
\beq \label{A5}
\zeta = \frac{1}{\sqrt{2}}\,(X + i\, Y)\,, \qquad \mathcal{W} =  \frac{1}{\sqrt{2}}\,(\mathcal{W}_1 + i\, \mathcal{W}_2)\,, 
\eeq
in Eq.~\eqref{A4}, we recover the standard form of the metric of the Kundt's class~\cite{R1, Kundt, KuTr}
\beq \label{A6}
ds^2 = -2\,du\,(dv + \mathcal{W}\,d\zeta + \bar{\mathcal{W}}\,d\bar{\zeta} + H\,du) +2\,P^{-2} \,d\zeta\,d\bar{\zeta}\,.
\eeq
It therefore remains to put Eq.~\eqref{A4} into the form of Eq.~\eqref{A3}. To this end, consider the coordinate transformation 
\beq \label{A7}
(u, v, X, Y) \mapsto (u, v, x, y)\,, \qquad  X = \frac{1}{4}\,\lambda\,\Psi^4(u, x)\,, \qquad Y = y\,,
\eeq
which turns Eq.~\eqref{A4} into Eq.~\eqref{A3} provided we have 
\beq \label{A8}
H = - \frac{1}{2}\,\lambda^2\,\Psi^4\,\Psi_u^2\,, \quad \mathcal{W}_1 = -\frac{4v}{\lambda\,\Psi^4} + \lambda\,\Psi\,\Psi_u\,, \quad \mathcal{W}_2 = 0\,, \quad P = \Psi\,. 
\eeq
Thus $\mathcal{W} = \mathcal{W}_1/\sqrt{2}$ is real for TGWs. 

For the special case of TGWs of Petrov type \emph{D}, we have $\Psi = u\, f(x) + h(x)$, where $f(x)$ and $h(x)$ are smooth integration functions, see the discussion following Eq.~\eqref{S2}.  The standard Kundt form of the metric of a particular class of solutions of Petrov type \emph{D} is given in Eq. (18.43) of Section 18.6 of Ref.~\cite{Griffiths:2009dfa}. It is useful to compare our general result given in Eq.~\eqref{A8} with Eq. (18.43). There is agreement if in our  $\Psi = u\, f(x) + h(x)$, we assume that  $f(x) = 0$, so that $\Psi_u = 0$, and the other quantities in Eq. (18.43) are given by $z = \lambda\,\Psi^2/2$, $n = \lambda/4$ and $\epsilon_0 = \epsilon_2 = \gamma = e = g = \Lambda = 0$~\cite{Griffiths:2009dfa}. We note that this solution is simply the static Kasner metric given by Eq.~\eqref{S6}.

\section{Weyl Curvature}

\subsection{Decomposition of the Weyl tensor in (1+3) and (1+1+2) parts}

The (1+3) threading decomposition of a tensor relative to an observer 4-velocity $U^\mu$ relies on the projection to the observer's rest space defined by
\begin{equation}\label{B1}
   P^{\mu}{}_{\nu} = \delta^{\mu}{}_{\nu} + U^\mu U_\nu\,.
\end{equation}
The tensor $P_{\mu \nu}$ then also plays the role of the induced metric for the rest space of $U$. The Weyl tensor can be decomposed in gravitoelectric and gravitomagnetic parts  by the definitions (see, e.g., Ref.~\cite{R1})
\begin{equation}\label{B2}
   E_{\mu \nu} = C_{\mu \alpha \nu \beta}\, U^\alpha U^\beta\,,\qquad  B_{\mu \nu} = \dW_{\mu \alpha \nu \beta}\, U^\alpha U^\beta\,,
\end{equation}
where $\dW_{\mu \nu \rho \sigma}$ is the dual of the Weyl curvature tensor. The tensors $E_{\mu \nu}$ and $B_{\mu \nu}$ are spatial, symmetric and traceless, each having five independent components. They can be further decomposed relative to a unit spacelike vector $n^\mu$ in the local rest space of $U$, $n\cdot U=0$. In a TGW spacetime, for instance, $n^\mu$ could be a component of the spatial triad of the fiducial observer indicating the direction of wave propagation. There is then a corresponding projection defined by
\begin{equation}\label{B3}
   \Phi^{\mu}{}_{\nu} = P^{\mu}{}_{\nu} - n^{\mu} n_{\nu}\,.
\end{equation}
Then, $\Phi_{\mu \nu}$ is a metric for the spatial 2-space (screen space) perpendicular to $U^\mu$ and $n^\mu$.
Given any spatial, symmetric and traceless tensor $\mathcal{S}_{\mu \nu}$, we define~\cite{Clifton:2016mxx}
\begin{equation}\label{B4}
 \begin{split}
        \fscalar{\mathcal{S}} &= \mathcal{S}_{\mu \nu} n^\mu n^\nu\,, \\
         \fvec{\mathcal{S}}^\mu &= \Phi^{\mu \nu} n^\rho \mathcal{S}_{\nu \rho}\,, \\
   \ftensor{\mathcal{S}}_{\mu \nu} &= (\Phi_{(\mu}{}^\rho{}_{\nu)}{}^\sigma - \tfrac12 \Phi_{\mu \nu}{}^{\rho \sigma}) \mathcal{S}_{\rho \sigma}\,,
 \end{split}
\end{equation}
where $\Phi_{\mu \nu \rho \sigma}= \Phi_{\mu \nu} \Phi_{\rho \sigma}$. Then $\mathcal{S}_{\mu \nu}$ can be written in terms of its parts as
\begin{equation}\label{B5}
   \mathcal{S}_{\mu \nu} = \fscalar{\mathcal{S}}_{\mu \nu} + \fvec{\mathcal{S}}_{\mu \nu} + \ftensor{\mathcal{S}}_{\mu \nu}
\end{equation}
where $\fscalar{\mathcal{S}}_{\mu \nu} =(n_\mu n_\nu - \frac12 \Phi_{\mu \nu}) \fscalar{\mathcal{S}}$ and 
$\fvec{\mathcal{S}}_{\mu \nu}= 2n_{(\mu} \fvec{\mathcal{S}}_{\nu)}$. 

\subsection{Weyl tensor for gravitational waves}
As is well known, localized gravitational energy is problematic in general relativity and does not have a covariant definition. However, the energy of a gravitational wave can in principle be measured. Even so, there is no general definition of the energy flux of gravitational waves. For a plane monochromatic wave of frequency $\omega$ propagating along the $z$ direction in linearized GR, the energy flux averaged over a wavelength is in the direction of wave propagation and is given by~\cite{Clifton:2014lha} 
\begin{equation}\label{B6}
   \frac{c^3\,\omega^2}{32\pi G}\,(h_{+}^2 + h_{\times}^2) =  \frac{c^3\,\omega^{-2}}{8\pi G}\,  <\mathcal{P}^z>\,,
\end{equation}
where $h_{+}$ and $h_{\times}$ represent  the constant amplitudes of the two independent linear polarization states of the wave and $<\mathcal{P}^z>$ is the corresponding average of the $z$ component of the  super-Poynting 4-vector $ \mathcal{P}_\mu$,
\begin{equation}\label{B7}
   \mathcal{P}_\mu = - e_{\mu \nu \rho} \quart E^{\nu}{}_{\sigma} \quart B^{\rho \sigma}\,.
\end{equation}
Here, $e_{\nu\rho\sigma} = U^\mu e_{\mu \nu \rho \sigma}$ is the spatial Levi-Civita tensor. Equation~\eqref{B6} is in agreement with the Landau-Lifshitz expression for the energy flux of the wave. 

\section{Gravity as an Optical Medium} 

The purpose of this appendix is to derive the constitutive relations when  $F_{\mu \nu} \to (\mathbf{E}, \mathbf{B})$ and  $\sqrt{-g}\,F^{\mu \nu} \to (-\mathbf{D}, \mathbf{H})$. This means that $F_{0i} = -E_i$ and $F_{ij} = \epsilon_{ijk}\,B_k$; similarly, $\sqrt{-g}\,F^{0i} = D_i$ and $\sqrt{-g}\,F^{ij} = \epsilon_{ijk}\,H_k$.  
Here, $\epsilon_{ijk}$ is the  totally antisymmetric symbol with $\epsilon_{123} = 1$. 
The derivation of the constitutive relations makes essential use of the fact that the matrices $(g_{\mu \nu})$ and $(g^{\mu \nu})$ are inverse of each other. The explicit expression of this fact, after separating temporal and spatial coordinates, results in four relations which can be used to show that 
$(\gamma_{ij})$ and its dual  $(\hat{\gamma}^{ij})$,
\begin{equation}\label{C1}
\gamma_{ij} := g_{ij} - \frac{g_{0i}\,g_{0j}}{g_{00}}\,, \qquad \hat{\gamma}^{ij} := g^{ij} - \frac{g^{0i}\,g^{0j}}{g^{00}}\,,          
\end{equation}
are inverses of $(g^{ij})$ and $(g_{ij})$, respectively; that is, $g^{ik}\,\gamma_{kj} = \delta^i_j$ and  
$g_{ik}\,\hat{\gamma}^{kj} = \delta_i^j$. The quantities in Eq.~\eqref{C1} also appear in the (1+3) timelike treading and the (3+1) spacelike slicing of spacetime into its temporal and spatial components~\cite{Bini:2012ht}; for instance, the spacetime metric adapted to the local environment of static observers with 4-velocity $U = (-g_{00})^{-1/2}\,\partial_t$ is given by
\begin{equation}\label{C1a}
ds^2 = g_{00}\,(dt - G_k\,dx^k)^2 + \gamma_{ij}\,dx^i\,dx^j\,.         
\end{equation}
The local space of the observers at rest is orthogonal to $U$, so that the corresponding induced metric is $P_{\mu \nu}\,dx^\mu\,dx^\nu =  \gamma_{ij}\,dx^i\,dx^j$, see Appendix B. 

Next, imagine computing the inverse of matrix $(g_{\mu \nu})$ using minors, etc. For the $g^{00}$ element of the inverse matrix and similarly for $g_{00}$, we find
\begin{equation}\label{C2}
g^{00} = \frac{\det(g_{ij})}{g}\,, \qquad g_{00} = g\,\det(g^{ij})\,,
\end{equation}        
where $g := \det(g_{\mu\nu})$. Extending this approach to the $g_{0k}$ element of the inverse of 
$(g^{\mu \nu})$ and similarly for $g^{0k}$, we find 
\begin{equation}\label{C3}
g \, \epsilon_{abc}\,g^{ai}\,g^{bj}\,g^{c0} = - \epsilon^{ijk}\,g_{0k}\,, \qquad  \frac{1}{g}\, \epsilon^{abc}\,g_{ai}\,g_{bj}\,g_{c0} = - \epsilon_{ijk}\,g^{0k}\,.
\end{equation}      
Finally, the definition of the determinant can be combined with Eq.~\eqref{C2} to show that 
\begin{equation}\label{C4}
\epsilon_{abc}\,g^{ai}\,g^{bj}\,g^{ck} = \epsilon^{ijk}\, \det(g^{pq}) = \epsilon^{ijk}\,\frac{g_{00}}{g}\,, \qquad  \epsilon^{abc}\,g_{ai}\,g_{bj}\,g_{ck} = \epsilon_{ijk}\,\det(g_{pq}) = \epsilon_{ijk}\,g\,g^{00}\,.
\end{equation}      

For the first constitutive relation, let us start with the \emph{electric} part of 
\begin{equation}\label{C5}
\sqrt{-g}\,F_{\mu \nu} = g_{\mu \alpha}\,g_{\nu \beta}\, \sqrt{-g}\,F^{\alpha \beta}\,,          
\end{equation}
which takes the form
\begin{equation}\label{C6}
- \sqrt{-g}\, E_j = g_{00}\,\gamma_{jk}\,D_k - g_{jp}\,g_{0q}\, \epsilon^{pqn}\, H_n\,.          
\end{equation}
Multiplying both sides of this relation by $g^{ij}/g_{00}$, then using $g^{ij}\,\gamma_{jk} = \delta^i_k$ and $g^{ij}\,g_{jp} = \delta^i_p - g^{i0}\,g_{0p}$, we get the first constitutive relation
$D_i = \epsilon_{ij}\,E_j - (\mathbf{G} \times \mathbf{H})_i$,         
where $\epsilon_{ij}$ and $\mathbf{G}$ are the dielectric tensor of the medium and its gyration vector, respectively, both given by Eq.~\eqref{P2}.  

For the second constitutive relation, we start with the \emph{magnetic} part of 
\begin{equation}\label{C7}
\sqrt{-g}\,F^{\mu \nu} = \sqrt{-g}\,g^{\mu \alpha}\,g^{\nu \beta}\, F_{\alpha \beta}\,,          
\end{equation}
so that we have
\begin{equation}\label{C8}
H_j = \frac{1}{2}\,\epsilon_{jmn}\,(\sqrt{-g}\, F^{mn}) =  \sqrt{-g}\,\epsilon_{jmn}\,[g^{mp}\,g^{0n}\,E_p+ \frac{1}{2}\,\epsilon_{abc}\,g^{ma}\,g^{nb}\,B_c]\,.          
\end{equation}
We now multiply both sides by $g^{ij}$ to get
\begin{equation}\label{C9}
g^{ij}\,H_j  =  \sqrt{-g}\,[\epsilon_{jmn}\,g^{ji}\,g^{mp}\,g^{n0}\,E_p+ \frac{1}{2}\,\epsilon_{abc}\,\epsilon_{jmn}\,g^{ji}\,g^{ma}\,g^{nb}\,B_c]\,.          
\end{equation}
Next,  using Eqs.~\eqref{C3} and~\eqref{C4}, we obtain the second constitutive relation 
$B_i = \mu_{ij}\,H_j + (\mathbf{G} \times \mathbf{E})_i$. 

The equality of the electric permittivity and magnetic permeability tensors , namely, 
\begin{equation}\label{C10}
\epsilon_{ij}  =  \mu_{ij} = -\sqrt{-g}\,\frac{g^{ij}}{g_{00}}\,          
\end{equation}
is a special property of the gravitational medium and implies the absence of \emph{double refraction}~\cite{VIS}.

\section{Simple Weak TGWs}

In Section IV, we use extensively the \emph{weak simple} TGWs given by Eq.~\eqref{S5} with $W = 1 + \phi$, where $\phi = \chi\,u + \eta\, x$, with arbitrary constant parameters $\chi$ and $\eta$, is such that $0 < |\phi| \ll 1$, see the discussion in Section IV following Eqs.~\eqref{M8} and~\eqref{M9}. To examine the uniqueness of this form of the metric for weak simple TGWs, we assume here that $W$ in the weak field solution of Eq.~\eqref{M8} is of the form
\beq \label{D1}
W = 1 + \varphi\,, \qquad   \varphi = u f(x) + h(x)-1\,, \qquad 0 < |\varphi| \ll 1\,,
\eeq
where $f$ and $h$ are arbitrary functions of $x$. It follows from Eq.~\eqref{M8} that we have a perturbation of Minkowski spacetime such that
\beq \label{D2}
g_{\mu \nu} = \eta_{\mu \nu} + h_{\mu \nu}\,, \qquad h_{\mu \nu} = \text{diag}(-4 \varphi, 4 \varphi, -2\varphi, 4\varphi).
\eeq
The perturbation away from Minkowski spacetime must satisfy linearized Einstein's equations, namely,
\beq \label{D3}
\Box h_{\mu \nu} - h_{\mu \alpha ,}{}^{\alpha}{}_{\nu} -  h_{\nu \alpha ,}{}^{\alpha}{}_{\mu}+ h_{\alpha}{}^{\alpha}{}_{,\,\mu \nu} + \eta_{\mu \nu}\,(h^{\alpha \beta}{}_{,\,\alpha \beta} - \Box h_{\alpha}{}^{\alpha})= 0\,.
\eeq
The result of substituting Eqs.~\eqref{D1} and~\eqref{D2} in Eq.~\eqref{D3} is that $f$ must be constant and $h(x)$ must depend linearly upon $x$. Thus $h_{\mu \nu}$ is a linear perturbation of Minkowski spacetime provided $\varphi = \chi \,u + \eta \,x$ up to an arbitrary constant that can be absorbed by a translation in $t$, $z$ or $x$. Here $\chi$ and $\eta$ are constants, as before, such that $0 < | \chi \,u + \eta \,x| \ll 1$.

The \emph{unique} perturbation is linear in spacetime coordinates; hence, the curvature tensor vanishes. We are thus in flat spacetime but in an accelerated coordinate system. Under a slight change of spacetime coordinates such that $x'^\mu = x^\mu - \epsilon^\mu$, we find 
\beq \label{D4}
h'_{\mu \nu} = h_{\mu \nu} + \epsilon_{\mu ,\, \nu} + \epsilon_{\nu , \,\mu}\,.
\eeq 
We can therefore change coordinates so that the metric is again the Minkowski spacetime metric, i.e. $h'_{\mu \nu} = 0$. It is straightforward to show that the corresponding coordinate transformation is given by 
\begin{eqnarray} \label{D5}
\epsilon^0 &=& -\chi [(t-z)^2 +x^2 -y^2/2] - 2 \eta t x\,,\nonumber\\
\epsilon^1 &=& -2\chi (t-z)x -\eta (t^2+x^2-z^2+y^2/2)\,,\nonumber\\
\epsilon^2 &=& \chi (t-z) y + \eta xy\,,\nonumber\\
\epsilon^3 &=& \chi [(t-z)^2-x^2+y^2/2] - 2 \eta x z\,.
\end{eqnarray}

\section*{Acknowledgments}

We are grateful to Georg Dautcourt for helpful correspondence. D.B. thanks the Italian INFN (Naples) for partial support.


\begin{thebibliography}{00}

\bibitem{Bini:2018gbq} 
  D.~Bini, C.~Chicone and B.~Mashhoon,
  ``Twisted gravitational waves'',
  Phys.\ Rev.\ D {\bf 97}, 064022 (2018)
  [arXiv:1801.06003 [gr-qc]].
 
 
 \bibitem{R1}
H.~Stephani, D.~Kramer, M.~MacCallum, C.~Hoenselaers and E.~Herlt, 
\emph{Exact Solutions of Einstein's Field Equations} (Cambridge University Press, Cambridge, UK,  2003), 2nd ed.

\bibitem{Daut1}
G.~Dautcourt, ``Zur Theorie der Gravitationsstrahlung", Ph.D. thesis, Humboldt University of Berlin (1962). 

\bibitem{Daut2}
G.~Dautcourt, ``Gravitationsfelder mit isotropem Killingvektor", in \emph{Relativistic Theories of Gravitation}, edited by L. Infeld (Pergamon Press, Oxford, UK, 1964), pp. 300-303.

  
  \bibitem{Harrison:1959zz} 
  B.~K.~Harrison,
   ``Exact three-variable solutions of the field equations of general relativity'',
  Phys.\ Rev.\  {\bf 116}, 1285 (1959).

\bibitem{DIRC}
R.~A.~d'Inverno and R.~A.~Russell-Clark, ``Classification of the Harrison metrics",
J.\ Math.\ Phys.\ {\bf 12}, 1258 (1971). 

\bibitem{Bini:2012ht} 
  D.~Bini, C.~Chicone and B.~Mashhoon,
  ``Spacetime splitting, admissible coordinates and causality'',
  Phys.\ Rev.\ D {\bf 85}, 104020 (2012)
  [arXiv:1203.3454 [gr-qc]].
  

\bibitem{Griffiths:2009dfa} 
  J.~B.~Griffiths and J.~Podolsk\'y,
  \emph{Exact Space-Times in Einstein's General Relativity} (Cambridge University Press, Cambridge, UK,  2009).
  
\bibitem{Mashhoon:1996wa} 
  B.~Mashhoon, J.~C.~McClune and H.~Quevedo,
  ``Gravitational superenergy tensor'',
  Phys.\ Lett.\ A {\bf 231}, 47 (1997)
  [gr-qc/9609018].

\bibitem{Mashhoon:1998tt} 
  B.~Mashhoon, J.~C.~McClune and H.~Quevedo,
  ``The gravitoelectromagnetic stress energy tensor'',
  Classical Quantum Gravity {\bf 16}, 1137 (1999)
  [gr-qc/9805093].  
  

\bibitem{bini-jan-min} 
D.~Bini, R.~T.~Jantzen and G.~Miniutti,
``Electromagnetic-like boost transformations of Weyl and minimal super-energy observers in black hole spacetimes",
 Int.\ J.\ Mod.\ Phys.\ D {\bf 11}, 1439-1450 (2002).
  
  
\bibitem{Clifton:2016mxx} 
 T.~Clifton, D.~Gregoris and K.~Rosquist,
  ``The magnetic part of the Weyl tensor, and the expansion of discrete universes'',
  Gen.\ Relativ.\ Gravit.\  {\bf 49},  30 (2017)
  [arXiv:1607.00775 [gr-qc]].
  

\bibitem{Bini:2017qnd} 
  D.~Bini, C.~Chicone and B.~Mashhoon,
  ``Anisotropic gravitational collapse and cosmic jets'',
  Phys.\ Rev.\ D {\bf 96}, 084034 (2017)
  [arXiv:1708.01040 [gr-qc]].
  
\bibitem{BoSe}
M.~\'A.~G.~Bonilla and J.~M.~M.~Senovilla,
"Very simple proof of the causal propagation of gravity in vacuum",
Phys.\ Rev. \ Lett.\ {\bf 78}, 783 (1997).
  
  
\bibitem{Sk}
G.~V.~Skrotskii, ``The influence of gravitation on the propagation of light", 
Sov. Phys. Doklady {\bf 2} 226-229 (1957). 

\bibitem{Pl}
J.~Plebanski, ``Electromagnetic waves in gravitational fields", 
Phys. Rev. {\bf 118}, 1396-1408 (1960). 

\bibitem{Fe}
F.~de Felice, ``On the gravitational field acting as an optical medium", 
Gen. Relativ. Gravit. {\bf 2}, 347-357 (1971). 

\bibitem{BM1}
B.~Mashhoon, ``Scattering of electromagnetic radiation from a black hole", 
Phys. Rev. D {\bf 7}, 2807-2814 (1973).

\bibitem{BM2}
B.~Mashhoon, ``Electromagnetic scattering from a black hole and the glory effect", 
Phys. Rev. D {\bf 10}, 1059-1063 (1974).

\bibitem{BM3}
B.~Mashhoon, ``Influence of gravitation on the propagation of electromagnetic radiation", 
Phys. Rev. D {\bf 11}, 2679-2684 (1975).

\bibitem{MaGr}
B.~Mashhoon and L.~P.~Grishchuk, ``On the detection of a stochastic background of gravitational radiation by the Doppler tracking of spacecraft", 
Astrophys. J. {\bf 236}, 990-999 (1980). 


\bibitem{Kopeikin:2001dz} 
  S.~Kopeikin and B.~Mashhoon,
  ``Gravitomagnetic effects in the propagation of electromagnetic waves in variable gravitational fields of arbitrary moving and spinning bodies'',
  Phys.\ Rev.\ D {\bf 65}, 064025 (2002)
  [gr-qc/0110101].


\bibitem{MK}
B.~Mashhoon and H.~Kaiser, ``Inertia of intrinsic spin",
 Physica  B  {\bf 385-386}, 1381 (2006)
[arXiv: quant-ph/0508182].

\bibitem{DSH}
B.~Demirel, S.~Sponar and Y.~Hasegawa, ``Measurement of the spin-rotation coupling in neutron polarimetry", 
New\ J.  Phys. {\bf 17}, 023065 (2015).

\bibitem{SRC}
D.~Kobayashi, T.~Yoshikawa, M.~Matsuo, R.~Iguchi, S.~Maekawa, E.~Saitoh and Y.~Nozaki,
``Spin current generation using a surface acoustic wave generated via spin-rotation coupling",
Phys.\ Rev.\ Lett.  {\bf 119}, 077202 (2017).


\bibitem{Mash}
B.~Mashhoon, ``Massless spinning test particles in a gravitational field", 
Ann.  Phys. (N.Y.)  \textbf{89}, 254 (1975). 

\bibitem{CMP}
C.~Chicone, B.~Mashhoon and B.~Punsly, ``Relativistic Motion of Spinning Particles in a Gravitational Field", Phys.\ Lett.  A \textbf{343}, 1-7 (2005)
[arXiv: gr-qc/ 0504146].

\bibitem{MS}
B.~Mashhoon and D.~Singh, ``Dynamics of Extended Spinning Masses in a Gravitational Field", 
Phys.\ Rev.\ D \textbf{74}, 124006 (2006)
[arXiv: astro-ph/0608278]. 

\bibitem{Bini:2017dny} 
  D.~Bini, A.~Geralico and A.~Ortolan,
  ``Deviation and precession effects in the field of a weak gravitational wave'',
  Phys.\ Rev.\ D {\bf 95}, 104044 (2017)
  [arXiv:1705.02794 [gr-qc]].
  
\bibitem{Collas:2018wcc} 
  P.~Collas and D.~Klein,
  ``Dirac particles in a gravitational shock wave'',
  Classical Quantum Gravity  {\bf 35}, 125006 (2018)
  [arXiv:1801.02756 [gr-qc]].
  
\bibitem{Kundt}
W.~Kundt, ``The plane-fronted gravitational waves", Z. Phys. {\bf 163}, 77-86 (1961).

\bibitem{KuTr}
W.~Kundt and M. Tr\"umper, ``Beitr\"age zur Theorie der Gravitations-Strahlungsfelder", Akad. Wiss. Lit. Mainz, Abhandl. Math.-Nat. Kl., No. 12 (1962); translated and reprinted in: Gen. Relativ. Gravit. {\bf 48}, 44 (2016).  
 
\bibitem{Clifton:2014lha} 
  T.~Clifton, D.~Gregoris and K.~Rosquist,
   ``Piecewise silence in discrete cosmological models'',
 Classical Quantum  Gravity  {\bf 31}, 105012 (2014)
[arXiv:1402.3201 [gr-qc]].



\bibitem{VIS}    
A.~M.~ Volkov, A.~A.~ Izmest'ev and. G.~V.~ Skrotskii, ``The propagation of electromagnetic waves in a Riemannian space", Sov.\ Phys.\ JETP {\bf 32}, 686 (1971). 



    

\end{thebibliography}
\end{document}